%% file: bhlflow.tex
\documentclass{aa}
\usepackage{amsmath}
\usepackage{amssymb}
\usepackage{graphicx}
\usepackage{latexsym}
\usepackage{aas_macros}
\usepackage{natbib}
\usepackage[squaren]{SIunits}


\input{myunits}

\newcommand{\vinf}[1]{v_{\infty}^{#1}}
\newcommand{\rhoinf}[1]{\rho_{\infty}^{#1}}

\newcommand{\zetaHL}[1]{\zeta_{\text{HL}}^{#1}}

\newcommand{\mdotHL}[1]{\dot{M}_{\text{HL}}^{#1}}

\newcommand{\racc}[1]{R_{\text{acc}}^{#1}}

\begin{document}

\title{High Mach-number Bondi--Hoyle--Lyttleton flow around a small accretor}

\author{Richard~G.~Edgar\inst{1}}

\offprints{Richard Edgar, \email{rge21@astro.su.se}}

\institute{Stockholms Observatorium, Albanova Universitetscentrum, SE-106 91, Stockholm, Sweden}

\date{Received / Accepted}

\abstract{In this paper, we discuss a two-dimensional numerical study of isothermal high Mach number Bondi--Hoyle--Lyttleton flow around a small accretor.
The flow is found to be unstable at high Mach numbers, with the instability appearing even for a larger accretor.
The instability appears to be the unstable radial mode of the accretion column predicted by earlier analytic work.
}

\titlerunning{High $\mathcal{M}$ BHL Flow around a small accretor}
\authorrunning{R.~G.~Edgar}

\maketitle


\input{intro}

\input{method}

\input{results}

\input{conclude}


\bibliographystyle{aa}
\bibliography{bibs/zeus,bibs/bondihoyle,bibs/reviews}

\begin{acknowledgements}
The author is supported by EU-RTN HPRN-CT-2002-00308, ``PLANETS'' and the calculations presented here were performed on the machines of HPC2N, Ume\aa{} and NSC, Link\"{o}ping.
Comments from an anonymous referee helped improve the discussion of the instability
\end{acknowledgements}

\end{document}

%% file: myunits.tex
\addunit{\cm}{\centi\metre}

\addunit{\dyne}{dyn}
\addunit{\erg}{erg}

\addunit{\gauss}{gauss}

\addunit{\persqcm}{\per \cm \squared}
\addunit{\persqcmnp}{\cm \rpsquared}

\addunit{\percubiccm}{\per \cm \cubed}
\addunit{\percubiccmnp}{\cm \rpcubed}

\addunit{\grampersqcm}{\gram \persqcm}
\addunit{\grampersqcmnp}{\gram \usk \persqcmnp}
\addunit{\grampercubiccm}{\gram \percubiccm}
\addunit{\grampercubiccmnp}{\gram \usk \percubiccmnp}

\addunit{\ergpercubiccm}{\erg \percubiccm}
\addunit{\ergpercubiccmnp}{\erg \usk \percubiccmnp}

\addunit{\molpercubiccm}{\mole \percubiccm}
\addunit{\molpercubiccmnp}{\mole \usk \percubiccmnp}

\addunit{\cmpersec}{\cm \per \second}
\addunit{\cmpersecnp}{\cm \usk \reciprocal \second}

\addunit{\cmpersecsq}{\cm \per \second \squared}
\addunit{\cmpersecsqnp}{\cm \usk \second \rpsquared}

\addunit{\gramcmpersec}{\gram \usk \cmpersec}
\addunit{\gramcmpersecnp}{\gram \usk \cmpersecnp}

\addunit{\gramsqcmpersec}{\gram \usk \cm \squared \per \second}
\addunit{\gramsqcmpersecnp}{\gram \usk \cm \squared \second \rpsquared}

\addunit{\yyear}{yr} 
\addunit{\MBU}{MBU} 

\addunit{\jansky}{Jy}

\addunit{\magnitude}{mag}

\addunit{\cmsqpergramnp}{\centi\metre\squared\usk\reciprocal\gram}

\newcommand{\mySun}{\odot}

\addunit{\Msol}{\ensuremath{\mathrm{M}_{\mySun}}}
\addunit{\Rsol}{\ensuremath{\mathrm{R}_{\mySun}}}
\addunit{\Lsol}{\ensuremath{\mathrm{L}_{\mySun}}}
\addunit{\Zsol}{\ensuremath{\mathrm{Z}_{\mySun}}}

\providecommand{\earth}{\oplus}

\addunit{\Mearth}{\ensuremath{\mathrm{M}_{\earth}}}
\addunit{\Rearth}{\ensuremath{\mathrm{R}_{\earth}}}

\addunit{\Mjup}{\ensuremath{\mathrm{M}_{J}}}
\addunit{\Rjup}{\ensuremath{\mathrm{R}_{J}}}

\addunit{\AU}{au}
\addunit{\lightyear}{ly}
\addunit{\parsec}{pc}

%% file: intro.tex
\section{Introduction}

The Bondi--Hoyle--Lyttleton (BHL) accretion problem concerns the accretion rate of a point mass moving at constant velocity through a uniform gas cloud.
This problem was first considered by \citet{1939PCPS.34..405}.
Considering Newtonian trajectories in a pressureless fluid, \citeauthor{1939PCPS.34..405} concluded that material with an impact parameter ($\zeta$) less than
\begin{equation}
\zeta < \zetaHL{} = \frac{2 G M}{\vinf{2}}
\label{eq:HoyleLyttletonRadiusDefine}
\end{equation}
would accrete onto the point mass, $M$ ($\vinf{}$ is the velocity of the gas relative to the point mass at infinity).
If the density of the gas is $\rhoinf{}$ (at infinity) then the accretion rate onto the point mass will be
\begin{equation}
\mdotHL{} = \pi \zetaHL{2} \vinf{} \rhoinf{} = \frac{4 \pi G^2 M^2 \rhoinf{}}{\vinf{3}}
\label{eq:HoyleLyttletonAccRateDefine}
\end{equation}
For a full discussion of the problem, see \citet{2004NewAR..48..843E}.
Numerical studies have shown (see \citet{2004NewAR..48..843E}, and references therein) that, although the accretion rate predicted by equation~\ref{eq:HoyleLyttletonAccRateDefine} is fairly accurate, the flow geometry is somewhat more complicated, due to the effects of pressure.

Bondi--Hoyle--Lyttleton accretion is of particular interest in star formation studies, since protostars in clusters are thought to undergo competitive accretion, at rates similar to that given by equation~\ref{eq:HoyleLyttletonAccRateDefine} \citep{2001MNRAS.323..785B}.
However, under such conditions, $\vinf{}$ is low, and so $\zetaHL{}$ is very large compared to the radius of the accretor ($\racc{}$).
This region of parameter space is poorly studied, due to the large computational expense involved.
\citet{2004MNRAS.349..678E} observed a flow instability while studing the effect of radiative feedback on Bondi--Hoyle--Lyttleton flow, and this paper looks more closely at this instability.
The computational method is discussed in section~\ref{sec:method} and the results in section~\ref{sec:results}.
Finally, the conclusions are presented in section~\ref{sec:conclude}.

%% file: method.tex
\section{Method}
\label{sec:method}

The {\sc Zeus-2d} code of \citet{1992ApJS...80..753S} was used to perform the calculations presented here.
The author parallelised the code using OpenMP, and found that it scaled well to up to eight processors.
Extra routines to monitor accretion rates and drag were added to the main code, and calculations were performed in spherical polar $(r, \theta)$ co-ordinates.

A linear grid of 100 cells was used in the $\theta$ direction, while the radial grid was logarithmically spaced.
The size of the innermost radial cells was chosen to make the inner portions of the grid as `square' as possible (to avoid possible truncation errors).
For the inner boundary condition, three alternatives were tried: {\sc Zeus}' own `outflow' boundary, a constant (small) density, and the density set to a small fraction of the innermost `active' cell.
The outer boundary lay at $10 \zetaHL{}$ (as defined by equation~\ref{eq:HoyleLyttletonRadiusDefine}) with a uniform inflow imposed on the upstream side, and outflow conditions on the downstream side.

The grid was initialised to a uniform flow, and the accretor introduced over a number of timesteps at the start of the run.
The simulations were run for as long as possible, to ensure that any `switch-on' effects died away.
The goal was a crossing time of the whole grid, but this was not possible for the runs with the small accretor.

In the simulations, dimensionless units were used.
The incoming material had $\vinf{}=1$, and the sound speed was set according to the required Mach number ($\mathcal{M}$).
The point mass was set to $M=1$ (as was the gravitational constant, $G$).
Using equation~\ref{eq:HoyleLyttletonRadiusDefine} we see that $\zetaHL{} = 2$ for these choices.
A value of $\rhoinf{}=0.01$ was chosen (arbitrarily, since the flow was isothermal and not self-gravitating), giving $\mdotHL{} = 0.126$ in these units.

%% file: results.tex
\section{Results}
\label{sec:results}

Three sets of simulations were performed.
For each of the sets, one run for each of the inner boundary conditions was made.
First, a low Mach number ($\mathcal{M}=3$) flow around a large ($\racc{} = 0.005 \zetaHL{}$) accretor was used to test the code.
The second set used the same size of accretor, but with the Mach number increased to $\mathcal{M}=20$.
For the final set, the accretor size was reduced to $0.0005 \zetaHL{}$.

We will now discuss the results of each of these sets in turn.


\subsection{Low Mach Number, Large Accretor}

\begin{figure}
\resizebox{\hsize}{!}{\includegraphics{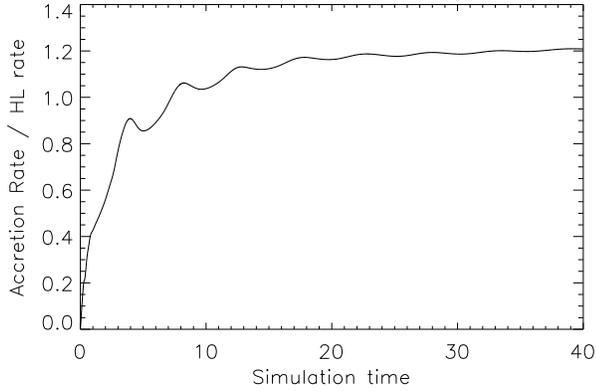}}
\caption{Accretion rates as a function of time for low Mach number flow around a large accretor}
\label{fig:LowLargeAccRates}
\end{figure}

\begin{figure}
\resizebox{\hsize}{!}{\includegraphics{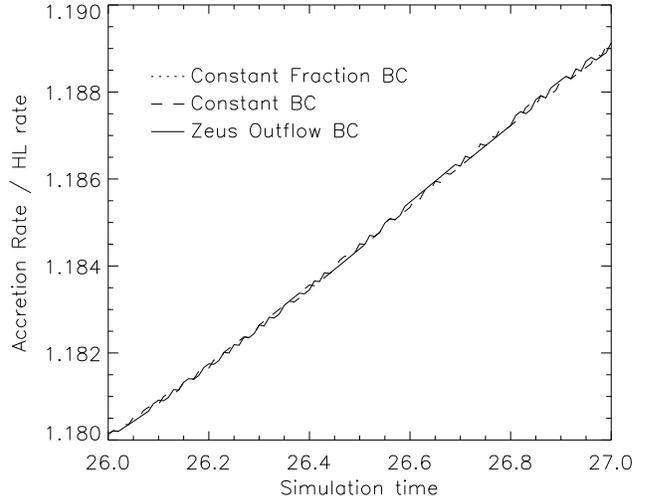}}
\caption{Detail of figure~\ref{fig:LowLargeAccRates} to show differences due to boundary conditions}
\label{fig:LowLargeAccRatesDetail}
\end{figure}

The accretion rate as a function of time is presented in figure~\ref{fig:LowLargeAccRates}.
After an initial turn-on period, the accretion rate settles to a fairly steady value, slightly larger than that predicted by equation~\ref{eq:HoyleLyttletonAccRateDefine}.
Changing the inner boundary condition does not affect the results significantly, as is shown in figure~\ref{fig:LowLargeAccRatesDetail}.
The different boundary conditions are giving slightly different curves, but are very close together.

\begin{figure}
\resizebox{\hsize}{!}{\includegraphics{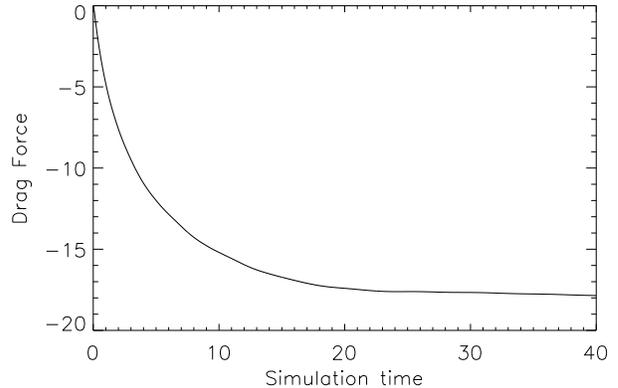}}
\caption{Gravitational drag force in units of $\mdotHL{} \vinf{}$ for low Mach number flow around a large accretor}
\label{fig:LowLargeDrag}
\end{figure}

The gravitational drag force on the accretor for this set of runs is plotted in figure~\ref{fig:LowLargeDrag}.
This is plotted in units of $\mdotHL{} \vinf{}$ which is the natural scaling for the problem.
The drag force rises smoothly to around $18 \mdotHL{} \vinf{}$.

Inspection of the density field shows a wake with a fairly broad opening angle, but no bow shock -  as is typical for isothermal BHL flow.
None of these results are particularly surprising, when compared to the work of \citet{1996A&A...311..817R} (model `GS' is closest to that presented here).
From this set of runs, we can conclude that the code is behaving well.


\subsection{High Mach Number, Large Accretor}

\begin{figure}
\resizebox{\hsize}{!}{\includegraphics[scale=0.5]{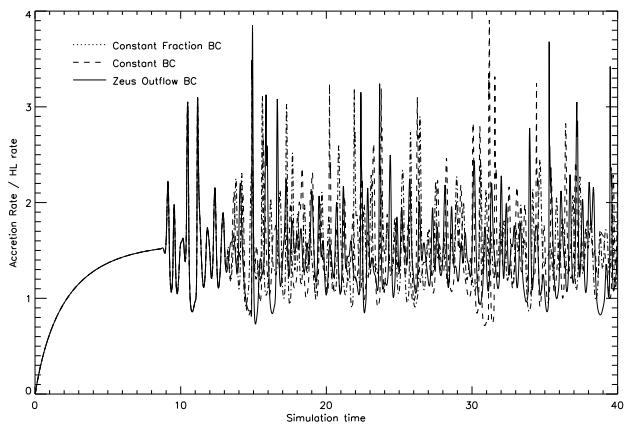}}
\caption{Accretion rates as a function of time for high Mach number flow around a large accretor}
\label{fig:HighLargeAccRates}
\end{figure}

The accretion rates for the three runs of high Mach number flow around a large accretor are presented in figure~\ref{fig:HighLargeAccRates}.
The behaviour is very different to that shown in figure~\ref{fig:LowLargeAccRates}, with a large oscillating accretion rate.
We can see that changing the boundary condition does change the flow history, but the change does not appear to be significant - the instability is similar in all cases.

\begin{figure}
\resizebox{\hsize}{!}{\includegraphics[scale=0.5]{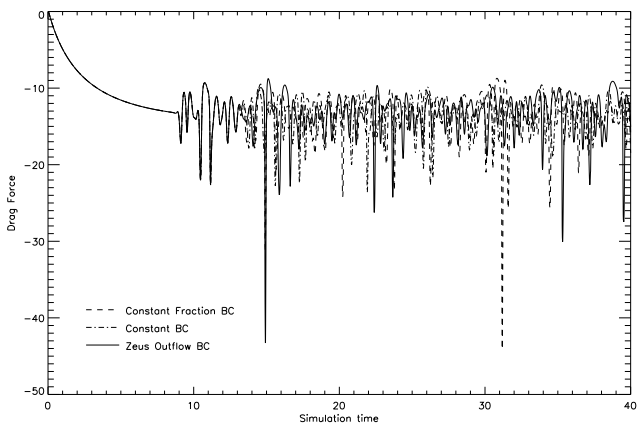}}
\caption{Gravitational drag (in units of $\mdotHL{} \vinf{}$) as a function of time for high Mach number flow around a large accretor}
\label{fig:HighLargeDrag}
\end{figure}

Figure~\ref{fig:HighLargeDrag} shows the corresponding gravitational drag for this set of runs.
As might be expected from the accretion rate curves, it is quite noisy.
Again, there are slight differences due to the inner boundary condition, but these do not seem to be significant compared to the general instability of the flow.

The wake's opening angle is much less than for the low Mach number case, as one would expect.
The oscillating accretion rate appears to be the `overstability' of radial oscillations described by \citet{1977MNRAS.180..491C} and \citet{1990ApJ...358..545S}.
The analysis was performed for a pressure-free (ballistic) flow, and the high Mach number isothermal nature of these calculations is a close approximation of this condition.
\citeauthor{1977MNRAS.180..491C} extended the analysis of \citet{1944MNRAS.104..273B} (see \citet{2004NewAR..48..843E} for a slightly more detailed exposition) to time dependent flow.
Perturbing the steady state solution and applying the WKB approximation (that is, the wavelength of the perturbation is assumed to be short compared to the scale on which it varies), \citeauthor{1977MNRAS.180..491C} found that the accretion column should be unstable.
\citeauthor{1990ApJ...358..545S} showed that viscosity could suppress this instability (which has not happened in the present work).
However, they disagreed on the physical nature of the instability: \citeauthor{1977MNRAS.180..491C} attributed it to the incoming material.
\citeauthor{1990ApJ...358..545S} asserted that the equation showing the possibilty for instability (the dispersion relation from the WKB approximation) did not contain terms relating to the inflowing material, and hence the instability was intrinsic to the flow along the line.
Settling this dispute is beyond the scope of the present work.

\begin{figure}
\resizebox{\hsize}{!}{\includegraphics[scale=0.5]{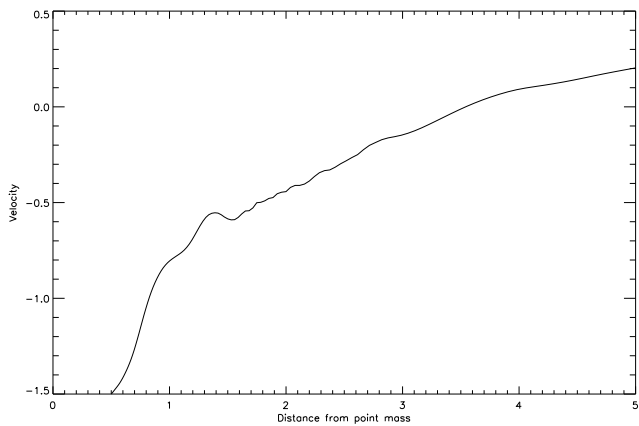}}
\caption{Sample radial velocity in the wake for high Mach number flow around a large accretor}
\label{fig:HighLargeSamplevr}
\end{figure}

Looking at a single accretion rate history curve from figure~\ref{fig:HighLargeAccRates}, around two to three oscillations occur per simulation time unit.
This is in accordance with the prediction of \citeauthor{1977MNRAS.180..491C}, who pointed out that the instability must grow faster than the characteristic accretion time ($\zetaHL{}/\vinf{}=2$ in this paper), or it will be swallowed by the accretor before it can grow.
Examining the velocity structure of the wake (figure~\ref{fig:HighLargeSamplevr} for a sample curve), oscillations in $v_r$ are observed superimposed on the freefall curve, as predicted by \citeauthor{1977MNRAS.180..491C}.
The oscillation wavelength does not vary much, as required by the WKB approximation.
The stagnation point is close to $r=3.5$, which is unsurprising, given that the accretion rates are slightly higher than $\mdotHL{}$ (which would place the stagnation point at $r=2$ -- cf \citet{2004NewAR..48..843E}).

\begin{figure}
\resizebox{\hsize}{!}{\includegraphics[scale=0.5]{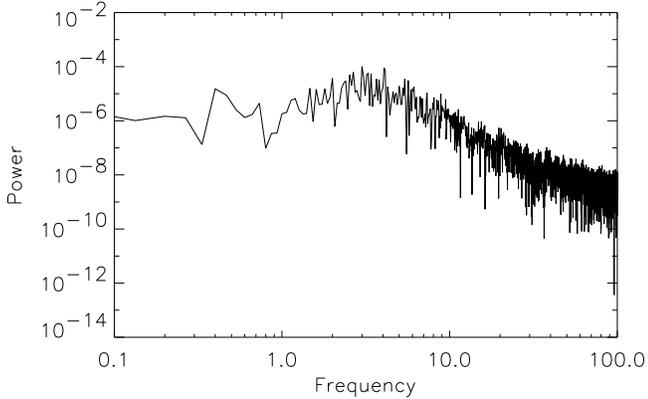}}
\caption{Sample accretion rate power spectrum for high Mach number flow around a large accretor}
\label{fig:HighLargeSamplePower}
\end{figure}

Unfortunately, computation of the power spectrum of the accretion rates (figure~\ref{fig:HighLargeSamplePower}) is rather unenlightening due to noise.
The spectrum peaks at a frequency of around $2$ to $3$ (it is too noisy to be more accurate), and has a power law fall off at higher frequencies.
There may also be a side peak at a frequency of around $0.5$, but this is only covered by a few points, and hence may just be noise.
The very intense bursts of accretion seen in figure~\ref{fig:HighLargeAccRates} seem to be associated with larger `blobs' of matter being accreted.
However, these are relatively rare, with only a couple appearing over the entire duration of the simulation (and one of those was probably a `switch-on' transient).

\subsection{High Mach Number, Small Accretor}

Finally, we come to the simulations of high Mach number flow around a small ($0.0005 \zetaHL{}$) accretor.
These simulations were painfully expensive computationally.
Each simulation in the first two sets required only about six days of wall time (on two processors).
These final three each occupied two processors for eight months, but still did not achieve the same elapsed simulation time.
As the grid is extended towards the point mass, the gravitational acceleration gets stronger.
There are more cells, the velocities are higher, and the cell size is smaller.
These three effects combine to cause the consumption of prodigious computing resources.

\begin{figure}
\resizebox{\hsize}{!}{\includegraphics[scale=0.5]{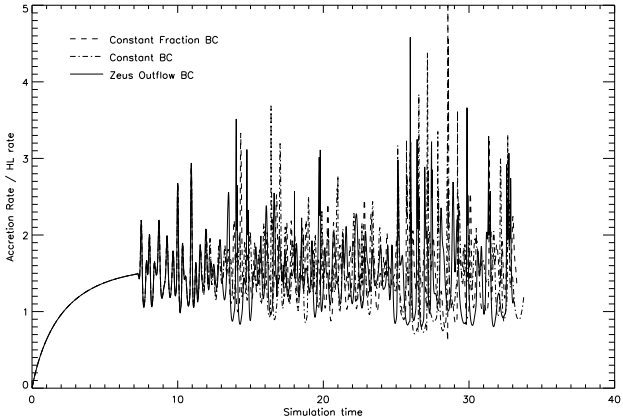}}
\caption{Accretion rates as a function of time for high Mach number flow around a small accretor}
\label{fig:HighSmallAccRates}
\end{figure}

\begin{figure}
\resizebox{\hsize}{!}{\includegraphics[scale=0.5]{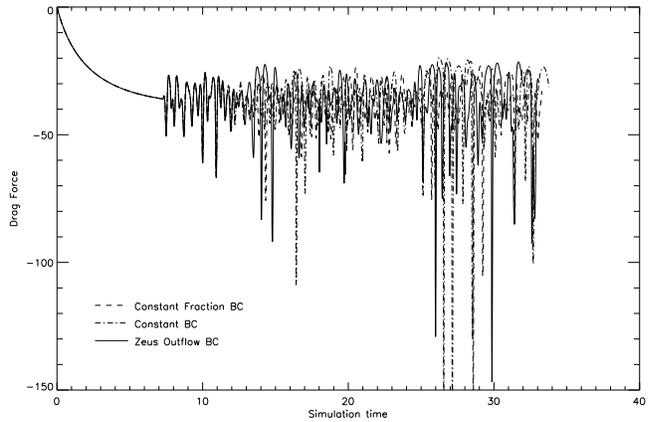}}
\caption{Gravitational drag (in units of $\mdotHL{} \vinf{}$) as a function of time for high Mach number flow around a small accretor}
\label{fig:HighSmallDrag}
\end{figure}

The accretion rate onto the point mass as a function of time is plotted in figure~\ref{fig:HighSmallAccRates}.
The behaviour is generally similar to figure~\ref{fig:HighLargeAccRates}, with large fluctuations once the flow becomes established.
There are differences in detail between the three boundary conditions, but the qualitative behaviour is obviously the same.
Figure~\ref{fig:HighSmallDrag} shows the gravitational drag for this set of runs.
Note that the gravitational force reaches larger values - as one might expect given the smaller size of the point mass (cf the Coulomb logarithm which appears in the calculation of dynamical friction. See e.g. \citet{1987gady.book.....B}).
If the flow is dense enough to make $\mdotHL{}$ interestingly large, then this drag force will alter $\vinf{}$ significantly.
However, the general pattern is similar to figure~\ref{fig:HighLargeDrag}, showing fluctuations comparable to those of the accretion rate.

The mean accretion rates for the two sets of simulations with high Mach number flow are presented in table~\ref{tbl:MeanAccRate}.
Shrinking the accretor does not seem to have made any significant difference to the results (beyond increasing the required computing resources by well over an order of magnitude).

\begin{table}
\caption{Mean Accretion Rates in units of $\mdotHL{}$ for high Mach number flow for different inner boundary conditions}
\label{tbl:MeanAccRate}
\centering
\begin{tabular}{cccc}
\hline \hline
Accretor Size   &   {\sc Zeus}  &  Const. $\rho$   &   Const. $\rho$ Frac. \\
\hline
Large  & 1.47  & 1.50  & 1.48 \\
Small  & 1.46  & 1.46  & 1.47 \\
\hline 
\end{tabular}
\end{table}

%% file: conclude.tex
\section{Conclusions}
\label{sec:conclude}

Based on the results presented above, we see that high Mach number Bondi--Hoyle--Lyttleton flow around a small accretor is unstable.
Although the average accretion rate is generally similar to that originally predicted by \citet{1939PCPS.34..405}, short increases to rates several times this are possible.
The instability seems to be related to the high Mach number of the flow - at lower Mach numbers, pressure must have a stabilising effect.
So far as these simulations can determine, the instability is the radial oscillation of the accretion column described by \citet{1977MNRAS.180..491C} and \citet{1990ApJ...358..545S}.

Observationally, the instability may be visible for deeply embedded young protostars.
For a \unit{1}{\Msol} protostar orbiting in a protocluster, the time unit in the present paper would correspond to approximately one year.
The fluctuations in the accretion rate would correspond to a brightening of a factor of a few over a period of days to weeks, and dimming over a similar timescale.
This is quite different from the FU Orionis phenomenon which is attributed to accretion disc activity (cf \citet{2000prpl.conf..897B} and references therein), where the outbursts rise on a timescale of months and last for years.

In the author's opinion, there is little to be gained by the further study of the pure Bondi--Hoyle--Lyttleton problem.
Shrinking the accretor size revealed no new behaviour, at a tremendous cost in computer time.
Instead, the Bondi--Hoyle--Lyttleton model should be retained as a useful `benchmark' and work focused on improving simulations of more realistic scenarios.